\documentclass[prd,superscriptaddress,amsfonts,amssymb,amsmath,showpacs,twocolumn]{revtex4-2}
\usepackage{bm}
\usepackage{amsfonts}
\usepackage{latexsym}
\usepackage[utf8]{inputenc}
\usepackage{graphicx}
\usepackage{amsmath}
\usepackage{palatino}
\usepackage{mathpazo}
\usepackage{textcomp}
\usepackage{float}
\usepackage{booktabs}
\usepackage{dcolumn}
\usepackage{ragged2e}
\usepackage{hyperref}
\hypersetup{colorlinks,citecolor=blue}
\hypersetup{colorlinks=true,linkcolor=red,filecolor=magenta,    urlcolor=blue}
\usepackage{amsmath}
\usepackage{xcolor}
\usepackage{orcidlink}
\usepackage[caption=false]{subfig}
\usepackage{commath}
\def\jnl@style{\it}
\def\aaref@jnl#1{{\jnl@style#1}}

\def\aaref@jnl#1{{\jnl@style#1}}

\def\aj{\aaref@jnl{AJ}}                   
\def\apj{\aaref@jnl{ApJ}}                 
\def\apjl{\aaref@jnl{ApJ}}                
\def\apjs{\aaref@jnl{ApJS}}               
\def\apss{\aaref@jnl{Ap\&SS}}             
\def\aap{\aaref@jnl{A\&A}}                
\def\aapr{\aaref@jnl{A\&A~Rev.}}          
\def\aaps{\aaref@jnl{A\&AS}}              
\def\mnras{\aaref@jnl{Mon.~Not.~Roy.~Astron.~Soc.}}             
\def\prd{\aaref@jnl{Phys.~Rev.~D}}        
\def\plb{\aaref@jnl{Phys.~Lett.~B}}        
\def\prc{\aaref@jnl{Phys.~Rev.~C}}  
\def\prl{\aaref@jnl{Phys.~Rev.~Lett.}}    
\def\qjras{\aaref@jnl{QJRAS}}             
\def\skytel{\aaref@jnl{S\&T}}             
\def\ssr{\aaref@jnl{Space~Sci.~Rev.}}     
\def\zap{\aaref@jnl{ZAp}}                 
\def\nat{\aaref@jnl{Nature}}              
\def\aplett{\aaref@jnl{Astrophys.~Lett.}} 
\def\apspr{\aaref@jnl{Astrophys.~Space~Phys.~Res.}} 
\def\physrep{\aaref@jnl{Phys.~Rep.}}      
\def\physscr{\aaref@jnl{Phys.~Scr}}       
\def\commat{\aaref@jnl{Comm.~Math.~Phys.}}              
\def\science{\aaref@jnl{Science}}               
\def\cqg{\aaref@jnl{Classical Quant.~Grav.}}            
\def\jpcs{\aaref@jnl{JPCS}}                                     
\def\ijmpd{\aaref@jnl{Int.~J.~Mod.~Phys.~D}}                    
\def\grg{\aaref@jnl{Gen.~Relat.~Gravit.}}               
\def\rpp{\aaref@jnl{Rep.~Prog.~Phys.}}          
\def\npa{\aaref@jnl{Nucl.~Phys.~A}}        
\def\lrr{\aaref@jnl{Living Rev.~Rel.}}                   
\def\jcap{\aaref@jnl{J.~Cosmology Astropart.~Phys.}}    
\def\rmp{\aaref@jnl{Rev.~Mod.~Phys.}}   
\def\epjc{\aaref@jnl{Eur.~Phys.~J.~C}}

\allowdisplaybreaks[1]
\renewcommand{\arraystretch}{1.1}
\begin{document}
\color{black}       
\title{Dark Energy Stars in Rastall-Rainbow Gravity: Structure, Stability and Observational Constraints}

\author{Ayan Banerjee \orcidlink{0000-0003-3422-8233}} 
\email{ayanbanerjeemath@gmail.com}
\affiliation{Astrophysics and Cosmology Research Unit, School of Mathematics, Statistics and Computer Science, University of KwaZulu--Natal, Private Bag X54001, Durban 4000, South Africa}

\author{Bobur Turimov\orcidlink{0000-0003-1502-2053}} 
\email{bturimov@astrin.uz}
\affiliation{Central Asian University, Milliy bog Str. 264, Tashkent, 111221, Uzbekistan}\affiliation{University of Tashkent for Applied Sciences, Str. Gavhar 1, Tashkent, 100149, Uzbekistan}

\author{Sulton Usanov}
\email{usanovsulton@gmail.com}
\affiliation{Kimyo International University in Tashkent, Usman Nasyr Str.156, Tashkent 100121, Uzbekistan}

\author{Murodbek Vapaev}
\email{murodbek.v@urdu.uz}
\affiliation{Deportment of Technique, Urgench State University, Kh. Alimjan Str. 14, Urgench 221100, Uzbekistan}

\author{Yunus Turaev}
\email{yunus.turaev.nw@gmail.com}
\affiliation{Mamun University, Bol-Khovuz Str. 2, Khiva 220900, Uzbekistan}

\author{Zebo Avezmuratova}
\email{zeboavezmuratova1981@mail.ru} 
\affiliation{Department of Physics and Asronomy, Urgench State Pedagogical Institute, Gurlan Str.1-A, Urgench 220100, Uzbekistan}


\date{\today}

\begin{abstract}
In this work, we investigate static configurations of dark energy stars within the framework of Rastall–Rainbow (R–R) gravity, which combines an energy-dependent deformation of spacetime with a nonminimal coupling between matter and geometry. We begin by deriving the modified field equations corresponding to R–R gravity and subsequently reformulate the stellar structure equations to describe hydrostatic equilibrium. The generalized Tolman–Oppenheimer–Volkoff (TOV) equations are then solved numerically by adopting the modified Chaplygin equation of state to model the interior matter distribution. The R-R parameters, along with fluid constants, are shown to influence the maximum mass, radii, and stiffness of the star sequences compared to the baseline set by general relativity. We apply observational benchmarks from high-mass pulsars and binary-merger events (e.g., GW170817 and GW190814) to appraise viability within the explored parameter space. The results collectively suggest that stable, causal configurations arise from physically meaningful parameter selections, with deviations from general relativity leading to systematic changes in structural characteristics while adhering to theoretical limits. 
These findings illustrate that Rastall-Rainbow gravity can support stable, observationally consistent dark energy stars, providing verifiable signatures in strong gravitational fields.

\end{abstract}

\maketitle

\section{Introduction}

Since its experimental confirmation during the 1919 solar eclipse \cite{Will:2014kxa}, Einstein's theory of general relativity (GR) has been a cornerstone of modern theoretical physics, withstood stringent tests at solar system scales, and is the basis for our understanding of gravitational phenomena in the cosmos. Its most remarkable predictions include the existence of compact astrophysical objects -black holes, neutron stars, and white dwarfs- which have transformed from mathematical demonstration to observational reality \cite{Ozel:2016oaf,Steiner:2017vmg}.

The discovery of pulsars in 1967 was the first direct evidence for the existence of neutron stars, which are extremely dense remnants of massive stars held together by neutron-degeneracy pressure. These extraordinary objects have masses usually around $M \sim 1$--$3\,M\odot$, 
which are compressed into roughly $10$--$15$ km radii, and have center densities exceeding nuclear saturation density  $\rho_{\mathrm{nuc}} = 2.8 \times 10^{14}\,\mathrm{g\,cm}^{-3}$. Since such extreme conditions cannot be reproduced in any terrestrial laboratory, the exact equation of state for matter at these densities remains one of the most interesting and unresolved problems in modern astrophysics and nuclear physics.

 Recent advances in observational astronomy have put increasingly strong constraints on the equation of state (EoS). NASA’s \textit{Neutron Star Interior Composition Explorer} (NICER) has given us precise mass–radius measurements via X-ray pulse-profile modeling \cite{Miller:2019cac}, while radio timing of millisecond pulsars has found very massive objects that rule out soft EoS: PSR~J0952$-$0607 at $2.35 \pm 0.17\,M_{\odot}$ \cite{Romani:2022jhd,Bassa:2017zpe}, PSR~J0740$+$6620 at $2.08 \pm 0.07\,M_{\odot}$ \cite{Fonseca:2021wxt}, and PSR~J0348$+$0432 at $2.01 \pm 0.04\,M_{\odot}$ \cite{Antoniadis:2013pzd}. These high-mass pulsars exclude a large part of the EoS that predicts maximum mass below $\sim 2\,M_{\odot}$. Meanwhile, gravitational-wave astronomy has opened a new window. The event GW170817 \cite{LIGOScientific:2018cki}—the first binary neutron-star merger observed in both gravitational waves (LIGO–Virgo) and across the electromagnetic spectrum—gave us model-independent constraints on the neutron-star tidal deformability $\Lambda$ and hence the EoS stiffness. The event GW190814 \cite{LIGOScientific:2020zkf} with a secondary component of $\sim 2.6\,M_{\odot}$ is either the lightest black hole or the heaviest neutron star ever seen, and sharpened the mass-gap puzzle. Recently, the source HESS~J1731$-$347 \cite{Doroshenko:2022nwp} has hinted at an ultra-light compact object with unusually low masses ($M \lesssim 0.8\,M_{\odot}$), raising the possibility of exotic matter configurations.

These observational developments motivate the exploration of alternative compact-object scenarios beyond conventional neutron stars. One intriguing possibility, originally proposed by Bodmer \cite{Bodmer:1971we}, Witten \cite{Witten:1984rs}, and Itoh \cite{Itoh:1970}, is that the ground state of bulk quantum chromodynamics (QCD) matter at sufficiently high density may consist of strange quark matter—a deconfined state containing roughly equal numbers of up, down, and strange quarks. Self-bound quark stars composed entirely or partially of strange quark matter can exhibit properties distinct from conventional neutron stars, including altered mass–radius relations and enhanced compactness \cite{Farhi:1984qu}. Another exotic alternative is the dark energy star, a hypothetical compact object supported by negative-pressure fluids obeying exotic equations of state. Such models, often based on Chaplygin-gas formulations \cite{Kamenshchik:2001cp, Bento:2002ps, Bilic:2001cg}, were originally motivated by cosmology to unify dark matter and dark energy behaviors \cite{Zhang:2004gc,Xu_2012} but have found applications in compact-object physics \cite{Pretel:2023nhf, Panotopoulos:2020kgl,Das:2023kmq,Pretel:2024tjw,Jyothilakshmi:2024zqn,Das:2024ugy,Banerjee:2025zhp}.

Whether these exotic configurations can actually exist depends on two things: the microphysics of matter at extreme densities, and the gravitational framework itself. While general relativity works remarkably well at solar-system scales, cosmological puzzles like accelerated expansion and dark matter hint that modifications might be needed at extreme curvatures or energies. This opens up interesting possibilities for exotic stellar structures that wouldn't survive in standard Einstein gravity.

Rastall gravity \cite{Rastall:1972swe} offers one such modification by allowing the energy-momentum tensor to couple directly with spacetime curvature through a non-conservation law. Although Visser \cite{Visser:2017gpz} showed that Rastall gravity can be rewritten as GR with an effective stress-energy tensor in vacuum, later work \cite{Darabi:2017coc,Oliveira:2015lka} clarified that real physical differences emerge when matter is present. Studies of compact stars \cite{Majeed:2023xde,ElHanafy:2022kjl,Ghosh:2021byh,Pretel:2024lae,Pattersons:2024fdz,Malik:2025ygg} have found that even small departures from GR can significantly shift maximum masses and stability boundaries. Physical differences from GR are also evident in black hole thermodynamics \cite{Lobo:2017dib,Laassiri:2024esn} and wormhole solutions \cite{ Moradpour:2016ubd,Mustafa:2019oiy,Saleem:2025jou,Halder:2019akt}

Rainbow gravity \cite{Magueijo:2002xx} takes a different approach, introducing energy-dependent spacetime metrics inspired by quantum gravity considerations \cite{ Assanioussi:2014xmz, Assanioussi:2016yxx}. The framework connects to doubly special relativity \cite{Gorji:2016laj, Kowalski-Glikman:2004fsz}, incorporating the Planck scale as an observer-independent energy scale \cite{ Amelino-Camelia:2003xax}. The idea is that particles with different energies effectively see different geometries. For compact stars with their extreme densities and strong fields, these modifications can substantially alter structural properties. Recent compact star studies \cite{ Hendi:2015vta, Tangphati:2025gnq, Rakhmanov:2025eup, BagheriTudeshki:2023dbm, Tudeshki:2022wed,Zhang:2020jmb,Zhang:2021fla} show that rainbow parameters can either enhance or suppress maximum masses depending on how the energy dependence is implemented. Physical effects of rainbow gravity are also manifest in black hole thermodynamics \cite{ Feng:2017gms, Li:2018gwf, Hendi:2015cra, Oubagha:2023vcc} and cosmological scenarios \cite{Hendi:2017cbu}.

Combining Rastall and rainbow frameworks into Rastall-Rainbow (R-R) gravity \cite{Mota:2019zln} creates a two-parameter theoretical space where both effects operate simultaneously. Previous R-R work has examined neutron stars \cite{Mota:2019zln}, wormholes 
\cite{Tangphati:2023nwz,Pradhan:2023vhn, Battista:2024gud,Kiroriwal:2025aum}, and various compact configurations \cite{Bora:2022qwe,Tangphati:2023fey} including Bose-Einstein condensate stars, which have also been investigated in \cite{Jyothilakshmi:2023cao}. However, dark energy stars in R-R gravity haven't been systematically studied yet, especially not with full comparisons to modern pulsar and gravitational-wave constraints.

We address this gap by investigating dark energy stars supported by the modified Chaplygin gas in R-R gravity. We solve the generalized Tolman-Oppenheimer-Volkoff equations for various combinations of Rastall and rainbow parameters, checking stability through three independent tests: static equilibrium conditions, adiabatic index profiles \cite{Chandrasekhar,Moustakidis:2016ndw}, and causality requirements. Our mass-radius curves are directly compared with observations of massive pulsars like PSR~J0952-0607 \cite{Romani:2022jhd}, PSR~J0740+6620 \cite{Fonseca:2021wxt}, and PSR~J0348+0432 \cite{Antoniadis:2013pzd}, plus constraints from gravitational-wave events GW170817 \cite{LIGOScientific:2018cki} and GW190814 \cite{LIGOScientific:2020zkf}, and the puzzling low-mass source HESS~J1731-347 \cite{Doroshenko:2022nwp}.

The paper proceeds as follows. Section~\ref{sec:field equations} derives the R-R field equations and reformulates them into a generalized stellar structure system. Section~\ref{sec:EoS} presents the modified Chaplygin gas equation of state. Section~\ref{sec:results} shows our numerical mass-radius sequences and observational comparisons. Section~\ref{sec:stability} analyzes stability for different parameter choices. Section~\ref{sec:conclusion} discusses implications and future directions.

 \section{Field Equations of Rastall-Rainbow Gravity}\label{sec:field equations}

  This work is established on two modified gravity theories: Rastall theory \cite{Rastall:1972swe} and rainbow gravity \cite{Magueijo:2002xx}, which offers an unique theoretical framework, identified as Rastall-Rainbow (R-R) gravity \cite{Mota:2019zln}, to study configurations of dark energy stars by means of energy-dependent spacetime geometry and non-minimal matter-geometry couplings.

\subsection{Rainbow Gravity Framework}

 Rainbow gravity emerges from extending doubly special relativity to curved spacetime, as originally proposed by Magueijo and Smolin~\cite{Magueijo:2002xx}. The central concept is that the geometry of spacetime becomes dependent on the energy of test particles moving through it. This energy-dependence changes completely the normal energy-momentum dispersion relation:

\begin{equation}
E^2\Xi^2(x) - p^2\Sigma^2(x) = m^2,
\end{equation}

\noindent where $x = E/E_p$ represents the dimensionless energy ratio between particle energy $E$ and Planck energy $E_p$. The functions $\Xi(x)$ and $\Sigma(x)$ serve as rainbow functions that characterize the energy-dependent modifications to spacetime.

These rainbow functions must satisfy specific boundary conditions to ensure physical consistency. In the low-energy limit, both functions approach unity:

\begin{equation}
\lim_{x \to 0} \Xi(x) = 1, \quad \lim_{x \to 0} \Sigma(x) = 1.
\end{equation}

This requirement guarantees that standard relativistic physics is recovered when particle energies are much smaller than the Planck scale.

The energy-dependent metric takes the form \cite{Magueijo:2002xx}:

\begin{equation}
g^{(x)}_{\mu\nu} = e^{(a)}_{\mu}(x) e^{(b)}_{\nu}(x) \eta_{ab},
\end{equation}

\noindent where the energy-dependent vierbein fields relate to energy-independent frame fields through:

\begin{equation}
e^{(0)}_{\mu}(x) = \frac{1}{\Xi(x)} e^{(0)}_{\mu}, \quad e^{(k)}_{\mu}(x) = \frac{1}{\Sigma(x)} e^{(k)}_{\mu},
\end{equation}

\noindent with $k = 1, 2, 3$ denoting spatial coordinates.

For a static, spherically symmetric spacetime appropriate for dark energy star configurations, this formalism yields the energy-modified metric:

\begin{equation}
ds^2 = -\frac{B(r)}{\Xi^2(x)}dt^2 + \frac{A(r)}{\Sigma^2(x)}dr^2 + \frac{r^2}{\Sigma^2(x)}d\Omega^2,
\end{equation}

\noindent where $d\Omega^2 = d\theta^2 + \sin^2\theta d\phi^2$ represents the standard angular element, and $A(r)$, $B(r)$ are metric potentials depending solely on the radial coordinate.

\subsection{Rastall Theory Formulation}

Rastall gravity represents a straightforward generalization of Einstein's general relativity, originally introduced in 1972~\cite{Rastall:1972swe}. The theory's fundamental premise challenges the standard conservation law in curved spacetime by proposing that energy-momentum tensor divergence couples directly to spacetime curvature.

Unlike general relativity where $\nabla_{\mu} T^{\mu\nu} = 0$, Rastall theory assumes \cite{Rastall:1972swe}:

\begin{equation}
\nabla_{\mu} T^{\mu\nu} = \bar{\lambda} R^{,\nu},
\end{equation}

\noindent where $\bar{\lambda}$ represents the coupling parameter that characterizes the strength of matter-geometry interaction.

This modified conservation equation can be rewritten as:

\begin{equation}
\nabla_{\mu} \left( T^{\mu\nu} - \frac{\bar{\lambda}}{4} \delta^{\mu\nu} R \right) = 0
\end{equation}

The corresponding field equations take the form:

\begin{equation}
R_{\mu\nu} - \frac{1}{2} g_{\mu\nu} R = 8\pi G \left( T_{\mu\nu} - \frac{\bar{\lambda}}{4} g_{\mu\nu} R \right).
\end{equation}

These can be rearranged into a more convenient form:

\begin{equation}
R_{\mu\nu} - \frac{\eta}{2} g_{\mu\nu} R = 8\pi G T_{\mu\nu},
\end{equation}

\noindent where $\eta = (1 + 2\bar{\lambda})^{-1}$ serves as the effective Rastall parameter. When $\eta = 1$, the standard Einstein field equations are recovered.

\subsection{Unified Rastall-Rainbow Theory}

The combination of these two theoretical frameworks creates the Rastall-Rainbow gravity theory \cite{Mota:2019zln} suitable for modeling dark energy stars. By incorporating energy-dependent metrics and gravitational constants into the Rastall formulation, we obtain the unified field equations:

\begin{equation}
R_{\mu\nu}(x) - \frac{\eta}{2} \delta_{\mu\nu}(x) R(x) = \kappa(x) T_{\mu\nu}(x),
\end{equation}

\noindent where $\kappa(x) = 8\pi G(x)$ and $G(x)$ represents the energy-dependent gravitational constant. Throughout this analysis, we adopt $G(x) = G$.

To facilitate practical calculations for dark energy star modeling, we reformulate these equations in Einstein-like form by introducing an effective energy-momentum tensor:

\begin{equation}
R_{\mu\nu} - \frac{1}{2} g_{\mu\nu} R = 8\pi \tilde{T}_{\mu\nu},
\end{equation}

\noindent where:

\begin{equation}
\tilde{T}_{\mu\nu} = T_{\mu\nu} - \frac{(1 - \eta)}{2\eta} g_{\mu\nu} T.
\end{equation}

For dark energy star investigations, we employ a perfect fluid energy-momentum tensor:

\begin{equation}
T_{\mu\nu} = (\rho + p)u_{\mu}u_{\nu} + p g_{\mu\nu}
\end{equation}

\noindent where $\rho$ denotes the dark energy density, $p$ represents the dark energy pressure, and $u^{\mu}$ is the four-velocity satisfying:

\begin{equation}
u^{\mu} = \left(\frac{\Xi(x)}{\sqrt{B(r)}}, 0, 0, 0\right).
\end{equation}

Applying this formalism to the energy-modified spherically symmetric metric yields the essential field equations governing dark energy star structure:

\begin{equation}\label{eq15}
M'(r) = 4\pi r^2 \tilde{\rho},
\end{equation}

\begin{equation}\label{eq16}
\frac{1}{r}\left(1 - \frac{2GM(r)}{r}\right)\frac{B'(r)}{B(r)} - \frac{2M(r)}{r^3} = 8\pi \tilde{p}.
\end{equation}

Here, we express the metric potential $A(r)$ in terms of the mass function $M(r)$ as $A(r)^{-1} = 1 - \frac{2M(r)}{r}$, and $\tilde{\rho}$ and $\tilde{p}$ represent the effective dark energy density and pressure:

\begin{equation}
\tilde{\rho} = \frac{1}{\Sigma^2(x)} \left[ \frac{2\eta - 1}{2(\eta - 1)} \rho + \frac{3(1 - \eta)}{2(\eta - 1)} p \right],
\end{equation}

\begin{equation}
\tilde{p} = \frac{1}{\Sigma^2(x)} \left[ \frac{2\eta - 1}{2(\eta - 1)} p + \frac{(1 - \eta)}{2(\eta - 1)} \rho \right] .
\end{equation}

The hydrostatic equilibrium equation for dark energy star configurations in R-R gravity becomes \cite{Mota:2019zln}:

\begin{equation}\label{eq19}
\frac{d\tilde{p}}{dr} = -\frac{(M + 4\pi \tilde{p}r^3)(\tilde{\rho} + \tilde{p})}{r^2(1 - 2M/r)}.
\end{equation}

This represents the modified Tolman-Oppenheimer-Volkoff equation that governs the equilibrium structure of dark energy stars within the Rastall-Rainbow framework. Together with Eqs.~(15), (16), and (19), along with an appropriate equation of state $\tilde{p} = \tilde{p}(\tilde{\rho})$ for dark energy matter, these equations form a complete system for determining the internal structure of dark energy stars in this modified gravity theory. In the following section, we present the specific equation of state that characterizes dark energy star configurations. 

\section{Chaplygin-inspired Dark Energy Model}
\label{sec:EoS}

\begin{figure*}[t]
    \centering
    \includegraphics[width=0.48\textwidth]{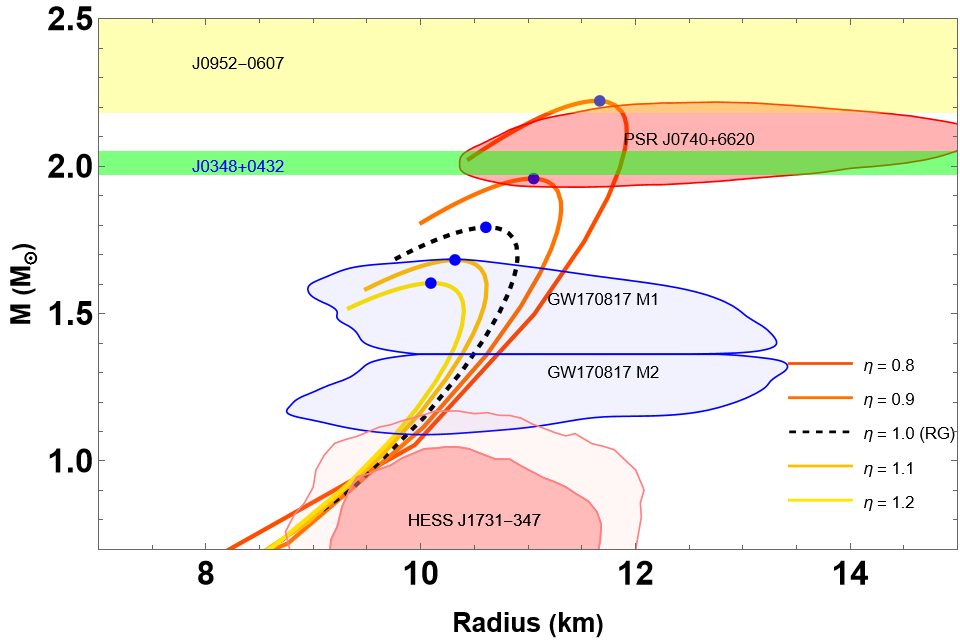}
    \includegraphics[width=0.48\textwidth]{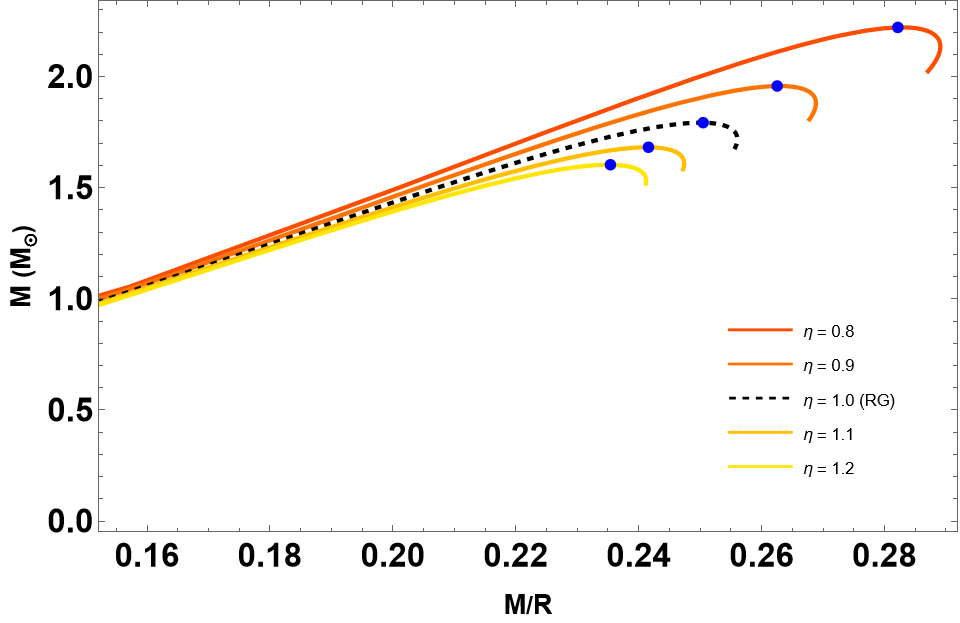}
    \caption{Mass-radius (left panel) and mass-compactness (right panel) relations for dark energy star configurations in Rastall-Rainbow gravity. The calculations are obtained with the parameter choice $G=1$, $A=\sqrt{0.4}$, $B = 0.23 \times 10^{-3}\,{\rm km}^{-2}$, and $\Sigma = 0.9$. The curves correspond to different values of the rainbow parameter $\eta = 0.8, 0.9, 1.0$ (Rainbow gravity limit), $1.1$, and $1.2$. Observational constraints are overlaid for PSR J0952-0607 \cite{Romani:2022jhd} (yellow band), PSR J0740+6620 \cite{Fonseca:2021wxt} (red contours), PSR J0348+0432 \cite{Antoniadis:2013pzd} (green band), and HESS J1731-347 \cite{Doroshenko:2022nwp} (magenta shaded region), together with the GW170817 \cite{LIGOScientific:2018cki} bounds (blue contours). Filled circles mark the maximum mass points along each sequence.}
    \label{fig1}
\end{figure*}

\begin{figure*}[t]
    \centering
    \includegraphics[width=0.48\textwidth]{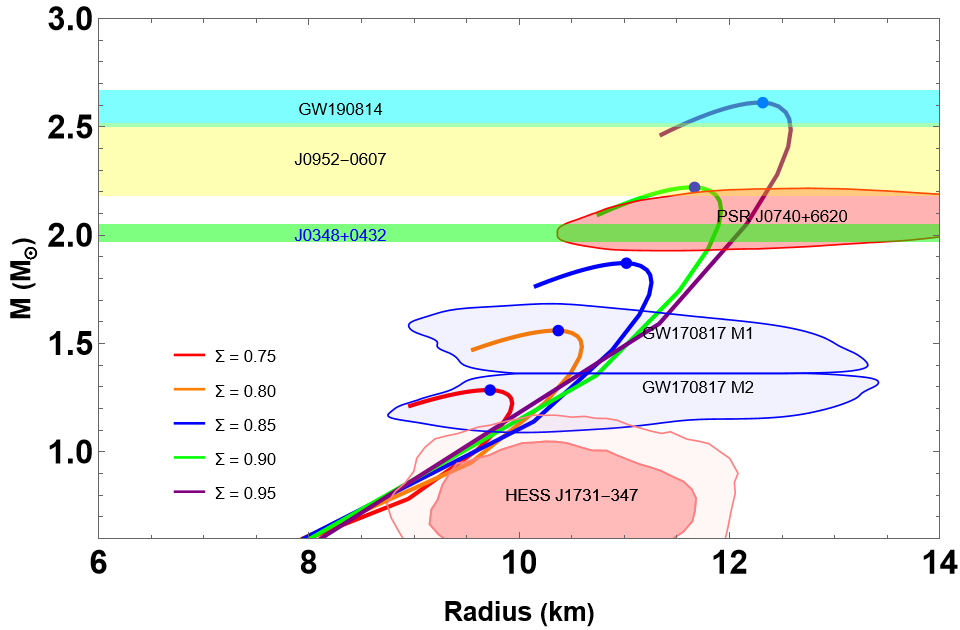}
     \includegraphics[width=0.48\textwidth]{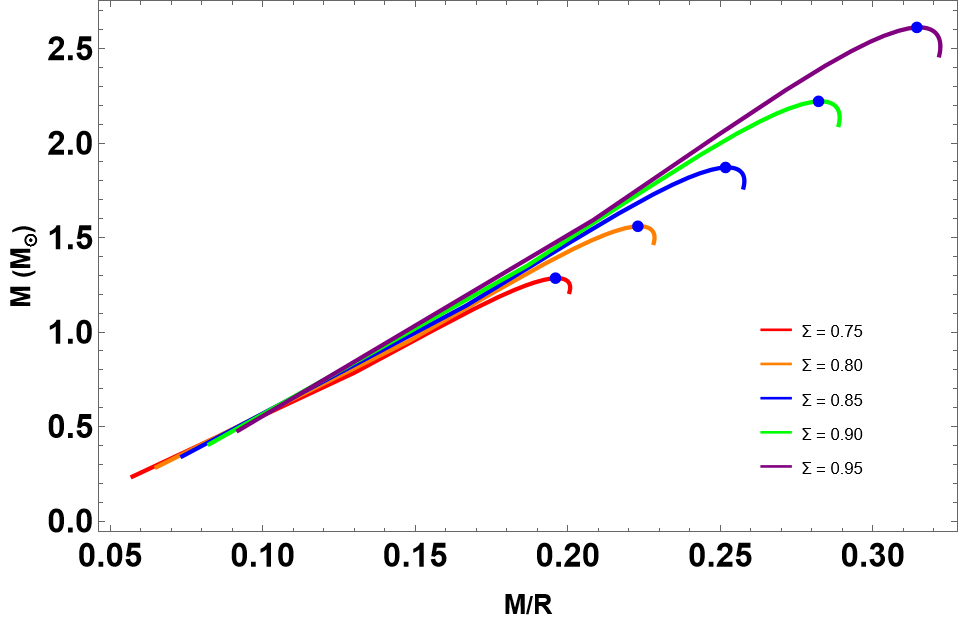}
    \caption{ Mass–radius (top panel) and mass–compactness (bottom panel) relations for dark-energy star configurations in Rastall–Rainbow gravity. The calculations use the parameter set \(G=1\), \(A=\sqrt{0.4}\), \(B=0.23\times10^{-3}\,\mathrm{km^{-2}}\), with fixed \(\eta=0.8\). The curves correspond to different values of the rainbow parameter \(\Sigma=\{0.75,0.80,0.85,0.90,0.95\}\). Observational constraints are overlaid for PSR~J0952–0607 \cite{Romani:2022jhd} (yellow band), PSR~J0740+6620 \cite{Fonseca:2021wxt} (red contours), PSR~J0348+0432 \cite{Antoniadis:2013pzd} (green band),  the GW190814 secondary mass range \cite{LIGOScientific:2020zkf} (cyan band), GW170817 component masses \cite{LIGOScientific:2018cki} (blue contours), and HESS~J1731–347 \cite{Doroshenko:2022nwp} (magenta shaded region). Filled circles mark the maximum-mass points along each sequence.}
    \label{fig2}
\end{figure*}

\begin{figure*}[t]
    \centering
    \includegraphics[width=0.48\textwidth]{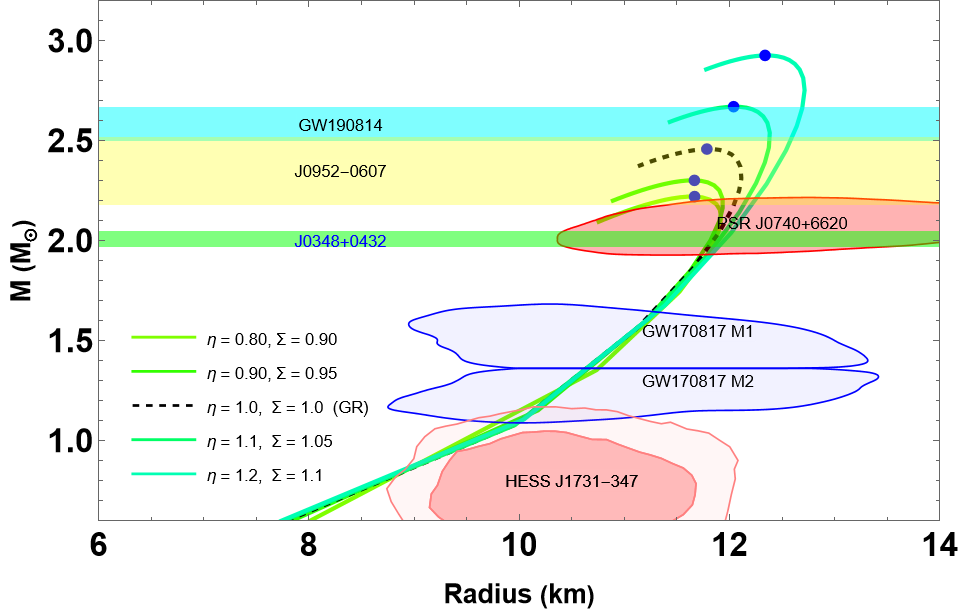}
     \includegraphics[width=0.48\textwidth]{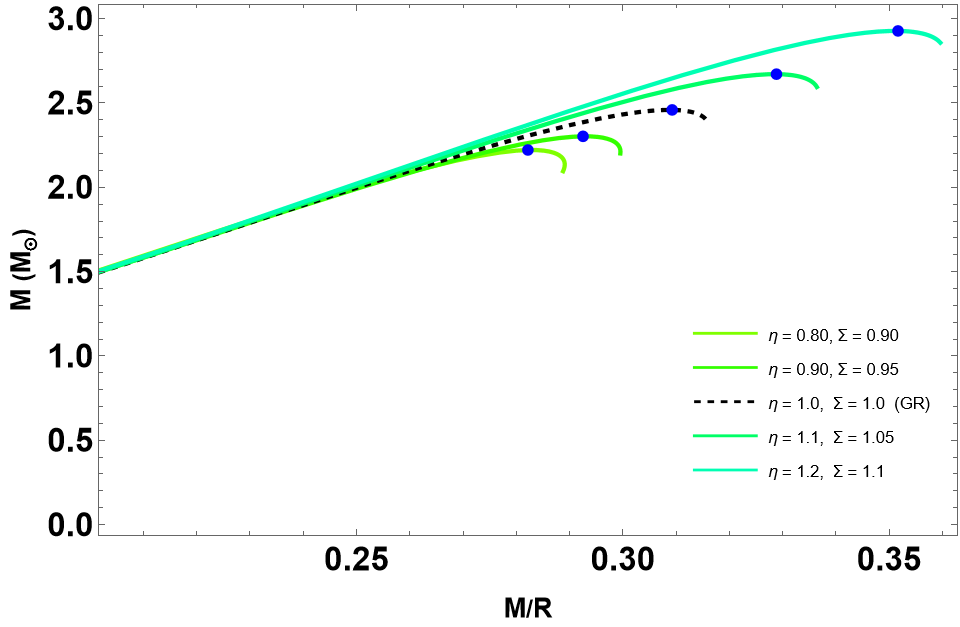}
    \caption{ Mass–radius (left panel) and mass–compactness (right panel) relations for dark-energy star configurations in Rastall–Rainbow gravity. Calculations use the parameter set \(G=1\), \(A=\sqrt{0.4}\), and \(B=0.23\times10^{-3}\,\mathrm{km^{-2}}\). Five sequences are shown for the pairs \((\eta,\Sigma)=\{(0.80,0.90),(0.90,0.95),(1.0,1.0),(1.1,1.05),(1.2,1.1)\}\); the black dashed curve denotes the GR baseline \((\eta=1,\Sigma=1)\). Observational constraints are overlaid for PSR~J0952--0607 \cite{Romani:2022jhd} (yellow band), PSR~J0740+6620 \cite{Fonseca:2021wxt} (red band), PSR~J0348+0432 \cite{Antoniadis:2013pzd} (green band), the GW190814 secondary-mass interval \cite{LIGOScientific:2020zkf} (cyan band), GW170817 component masses \cite{LIGOScientific:2018cki} (blue contours), and HESS~J1731--347 \cite{Doroshenko:2022nwp}(magenta shaded region). Filled circles mark the maximum-mass (radial stability turning-point) configurations on each sequence, and the same points are repeated in the compactness plane.
}
    \label{fig3}
\end{figure*}

 The Chaplygin gas scenario \cite{Kamenshchik:2001cp}, originally motivated by developments in string theory \cite{Ogawa:2000gj}, has long been considered as a candidate for describing exotic fluids in cosmology. Its most striking feature is the ability to unify dark matter and dark energy within a single fluid description \cite{Bento:2002ps,Bilic:2001cg,Gorini:2002kf,Zhang:2004gc,Xu_2012}. Despite this appealing property, recent observational constraints have challenged the viability of the original model. This has led to the proposal of extended formulations, notably the generalized Chaplygin gas \cite{Bilic:2001cg} and the modified Chaplygin gas (MCG) \cite{Debnath:2004cd,Saadat:2013ava}, which offer greater flexibility in fitting astrophysical and cosmological data. 

\par
For our analysis of compact stellar objects, we focus on the modified version of the Chaplygin prescription \cite{Pourhassan:2013sw,Saadat:2013ava}, whose EoS is given by \cite{Kahya:2015dpa}:
\begin{align}
p = A^2 \rho - \frac{B^2}{\rho},
\label{EOS}
\end{align}
where $A$ is a dimensionless parameter and $B$ carries units of energy density. Both constants are assumed to be positive. At the stellar boundary, where the  pressure vanishes, the surface energy density is determined by the relation $\rho_s = B/A$. The case $A=0$ corresponds to the conventional Chaplygin form. To remain consistent with causality, the model typically requires $A^2<0.5$, ensuring that the squared sound speed $v_s^2 = dp/d\rho$ remains subluminal \cite{Pretel:2023nhf}. In this work, we adopt these conditions to explore stellar interiors constructed from the MCG framework, as defined by Eq.~(\ref{EOS}) (see also Refs. \cite{Panotopoulos:2021dtu,Panotopoulos:2020kgl,Pretel:2023nhf}).


\section{Numerical results and discussion} \label{sec:results}

\begin{figure*}
	\centering
	\includegraphics[width=0.32\textwidth]{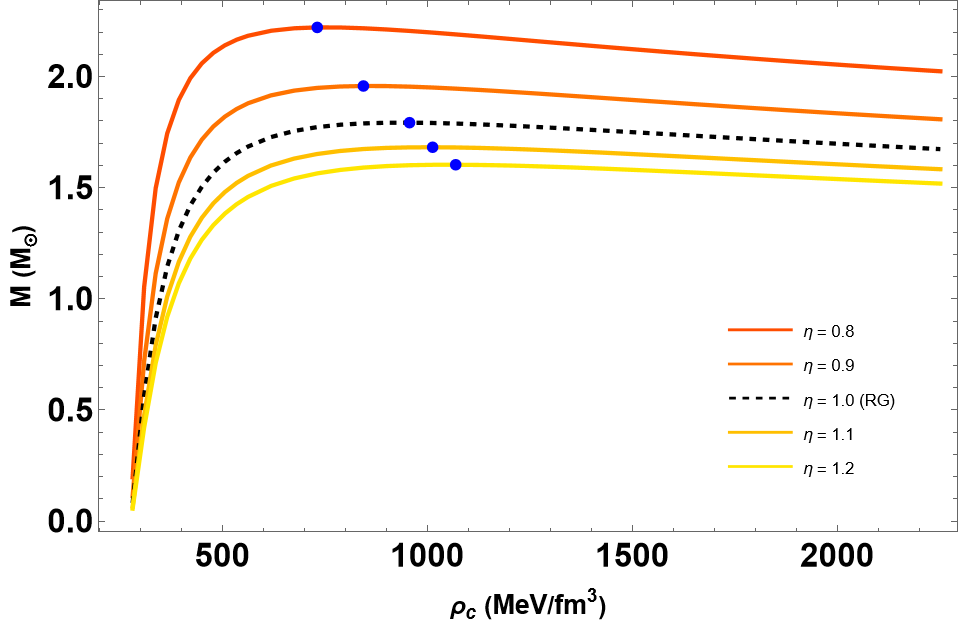}
	\includegraphics[width=0.32\textwidth]{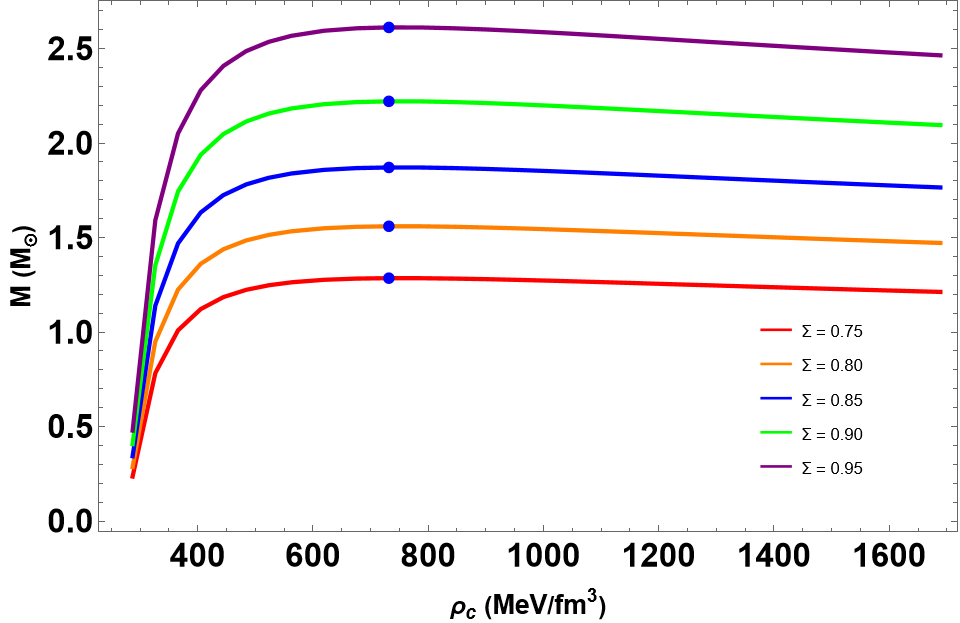}
	\includegraphics[width=0.32\textwidth]{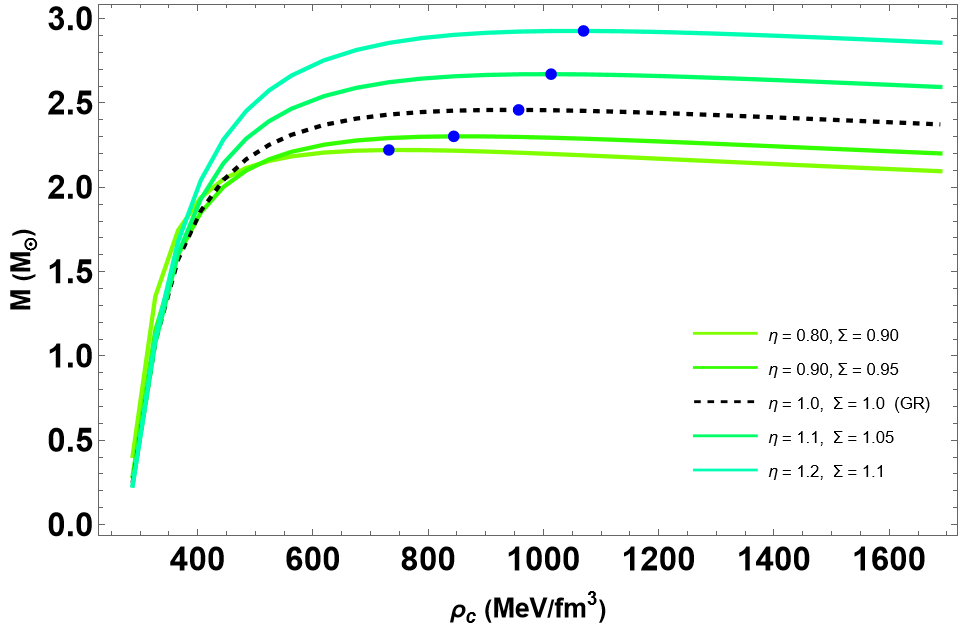}
	\caption{Profiles of the mass--central energy density relation $M(\rho_c)$ for dark energy stars in Rastall--Rainbow gravity. \textit{Left:} variation with the Rastall parameter $\eta$ (Table~\ref{table1}). \textit{Middle:} variation with the rainbow parameter $\Sigma$ (Table~\ref{table2}). \textit{Right:} joint variations around the GR baseline, $(\eta,\Sigma)$, as summarized in Table~\ref{table3}. The dashed black curve denotes the GR case $(\eta=1,\Sigma=1)$. Blue markers indicate the turning point (maximum mass) along each sequence; these configurations correspond to the entries reported in the respective tables. Unless stated otherwise, all remaining model constants are fixed as in the main text.}
	\label{fig4}
\end{figure*}

\begin{figure*}
	\centering
	\includegraphics[width=0.32\textwidth]{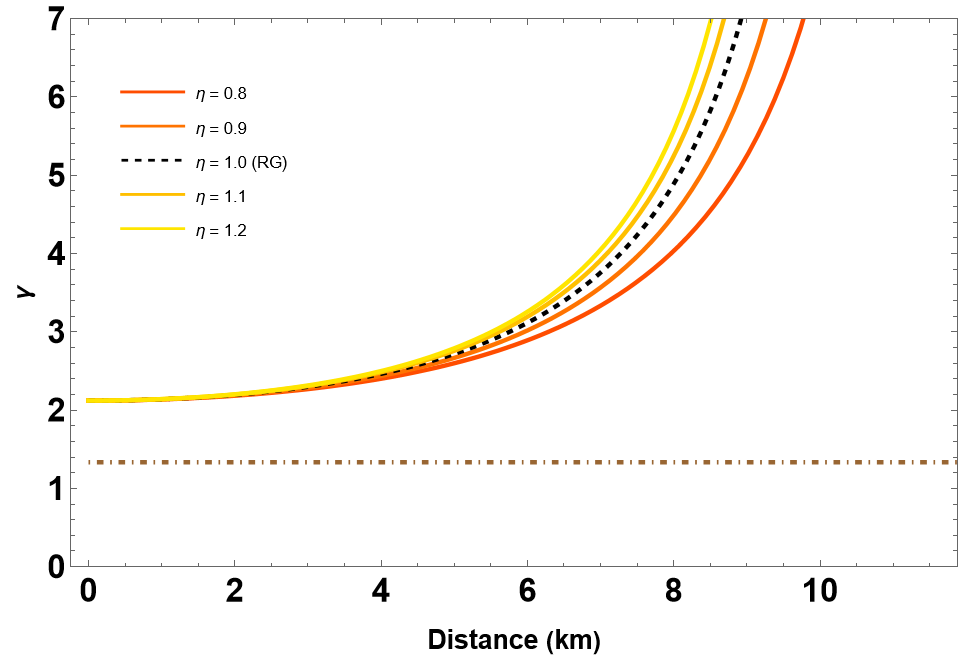}
	\includegraphics[width=0.32\textwidth]{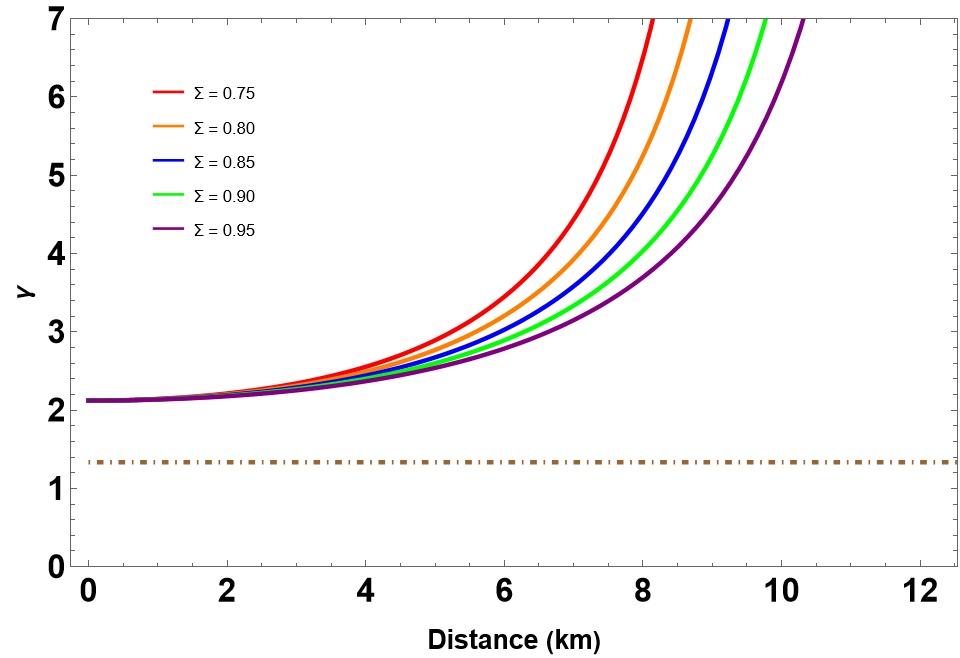}
	\includegraphics[width=0.32\textwidth]{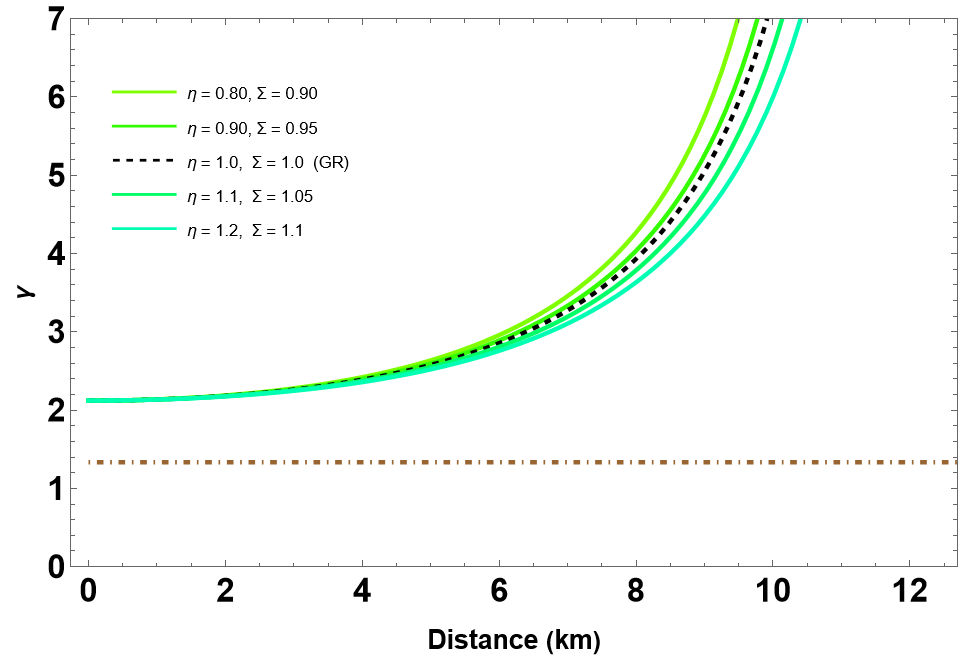}
	\caption{Variation of the adiabatic index $\gamma(r)$ for \textit{dark energy stars} in Rastall--Rainbow gravity. \textit{Left:} dependence on the Rastall parameter $\eta$ (Table~\ref{table1}). \textit{Middle:} dependence on the rainbow parameter $\Sigma$ (Table~\ref{table2}). \textit{Right:} joint variations around the GR baseline $(\eta,\Sigma)$, as summarized in Table~\ref{table3}. The horizontal dashed line marks the dynamical–stability threshold $\gamma=4/3$. For the parameter choices shown, $\gamma(r)$ remains above $4/3$ throughout the stellar interior, thereby satisfying the local stability criterion. Unless stated otherwise, all remaining model constants are fixed as in the main text.}
	\label{fig5}
\end{figure*}

\begin{figure*}
	\centering
	\includegraphics[width=0.32\textwidth]{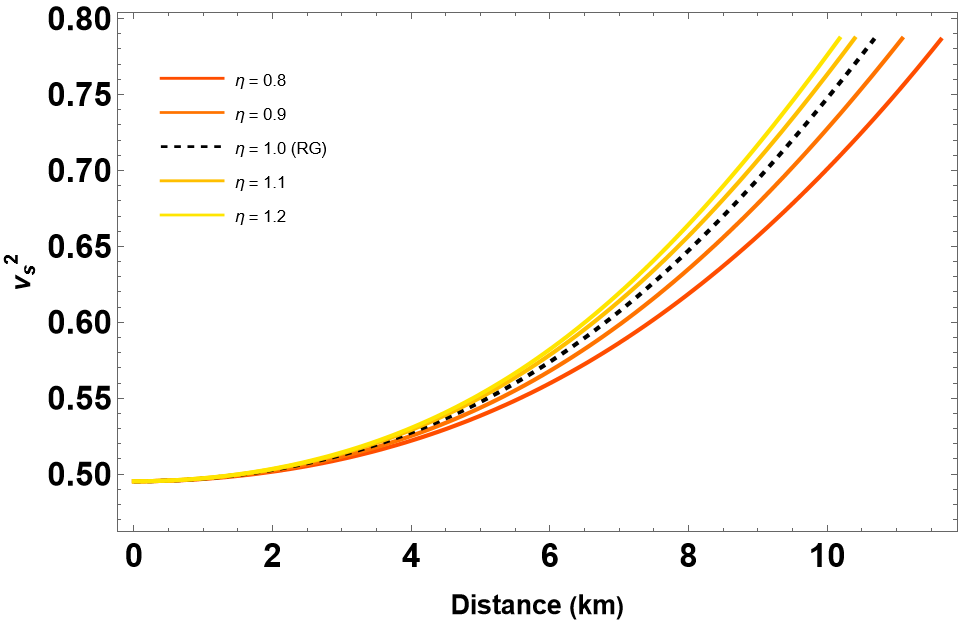}
	\includegraphics[width=0.32\textwidth]{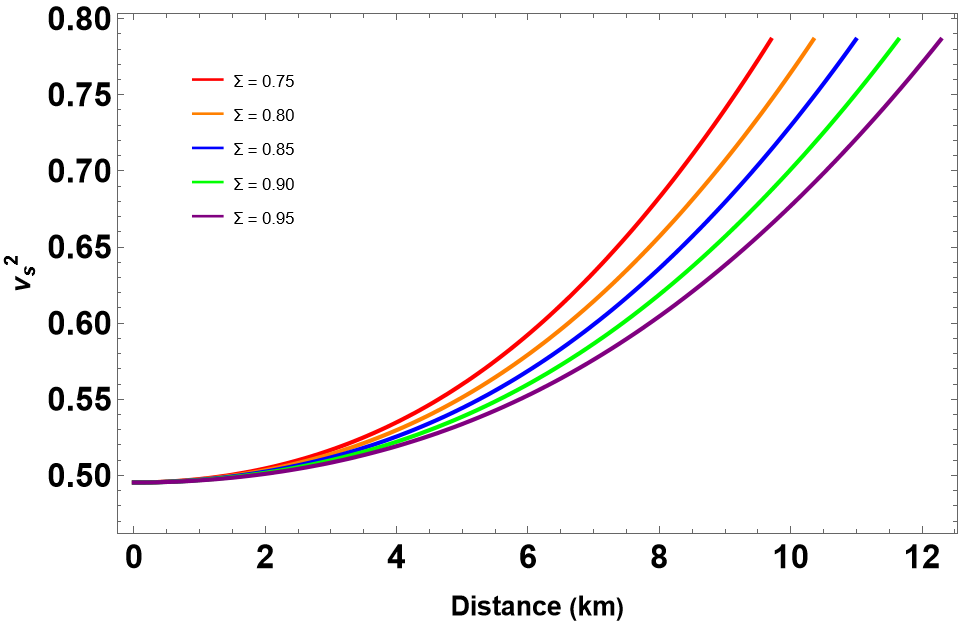}
	\includegraphics[width=0.32\textwidth]{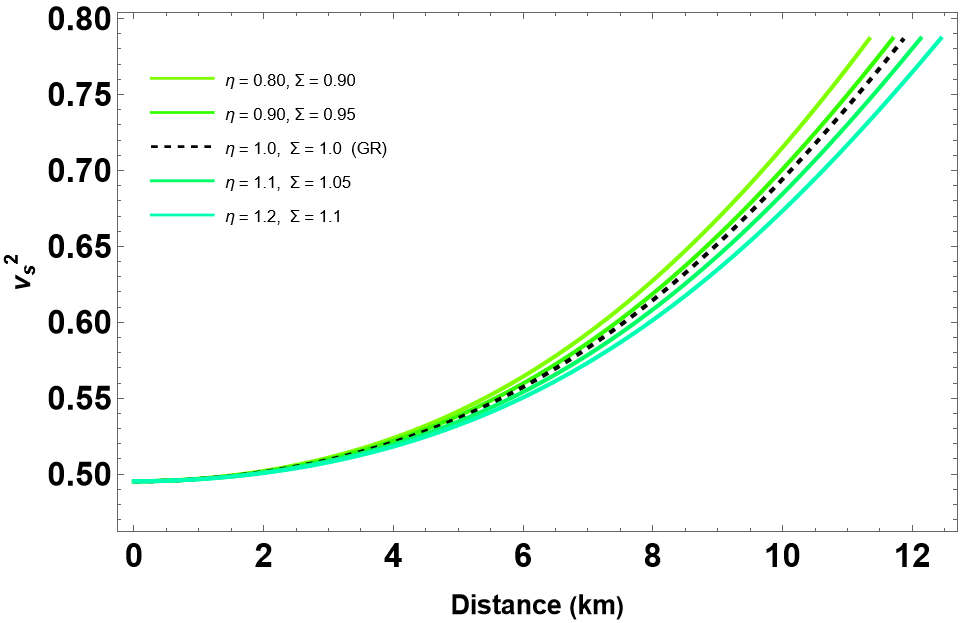}
\caption{Sound–speed profiles $c_{s}^{2}(r)$ for \textit{dark energy stars} in Rastall--Rainbow gravity. \textit{Left:} variation with the Rastall parameter $\eta$ (Table~\ref{table1}). \textit{Middle:} variation with the rainbow parameter $\Sigma$ (Table~\ref{table2}). \textit{Right:} joint variations around the GR baseline $(\eta,\Sigma)$, as summarized in Table~\ref{table3}. The dashed black curve denotes the GR case $(\eta=1,\Sigma=1)$. For all parameter choices shown, the causality condition $0\le c_{s}^{2}\le 1$ is satisfied throughout the stellar interior. Unless stated otherwise, all remaining model constants are fixed as in the main text.}
	\label{fig6}
\end{figure*}

 In this section, we solve numerically Eqs.~(\ref{eq15}) and (\ref{eq19}) to obtain the mass-radius profiles of dark energy stars and to investigate their internal physical properties.  The boundary conditions are established to ensure regularity at the center of the star \begin{eqnarray}
    m(r = 0) = 0, \text{ and } \rho(r = 0) = \rho_c. \label{inner_BC}
\end{eqnarray} 
where \(\rho_c\) denotes the central energy density. Then, by varying \(\rho_c\) appropriately, one obtains a sequence of stellar configurations with different masses and radii. The numerical integration proceeds outward from the center until the pressure reaches the stellar surface at which \(p(r=R)=0\)). Here, the stellar boundary defines by, \(R\).

Additionally, to ensure a smooth junction with the exterior spacetime, we match the interior solution at the stellar surface \(r=R\) (defined by \(p(R)=0\)) to the Schwarzschild vacuum. This imposes the boundary condition
\begin{equation}
B(R)=1-\frac{2\mathcal{M}}{R}, \label{eq:SchwMatch}
\end{equation}
where \(\mathcal{M}\equiv M(R)\) is the total gravitational mass of the star.  

\subsection{Profiles for variation of Rastall free parameter $\eta$}

\begin{table}
\caption{\label{table1} 
 Maximum mass configurations of dark energy stars in Rastall-Rainbow gravity for different values of the rainbow parameter $\eta$ with $\Sigma = 0.9$. Listed are the stellar mass $M$, corresponding radius $R_{M}$, central density $\rho_{c}$, and compactness $M/R$. The case $\eta=1.0$ represents the Rainbow gravity (RG) limit.}
\begin{ruledtabular}
\begin{tabular}{ccccc}
$\eta$  & $M$ [$M_\odot$]  &  $R_{M}$  [\rm{km}]  & $\rho_c$ [MeV/fm$^3$] & $M/R$  \\
\colrule
0.8  &  2.22  &  11.66  & 731 &  0.281  \\
0.9  &  1.95  &  11.05  & 844 &  0.260  \\
1.0 (RG) &  1.79  &  10.60  & 956 &  0.250  \\
1.1  &  1.68  &  10.31  & 1,013 &  0.241  \\
1.2  &  1.60  &  10.09  & 1,069 &  0.235
\end{tabular}
\end{ruledtabular}
\end{table}

 The findings shown in Table~\ref{table1} and Fig.~\ref{fig1} provide a systematic characterization of maximum mass configurations for dark energy stars under Rastall-Rainbow gravity as the Rastall parameter \(\eta\) varies. The maximum mass decreases monotonically from \(2.22\,M{\odot}\) to \(1.60\,M{\odot}\) while \(\eta \) increases from from \(0.8\) to \(1.2,\) and the corresponding radius shrinks from \(11.66\,km\) to \(10.09\,km\). In conjunction with this behavior, the central density shows a monotonic increase from \(731\,\mathrm{MeV\,fm^{-3}}\) when \(\eta=0.8\) to \(1069\,\mathrm{MeV\,fm^{-3}}\) when \(\eta=1.2\). The value of the compactness parameter shows a continuing decrease from \(M/R = 0.281\) to \(0.235\) for all configurations which remain safely under the Buchdahl bound of \((M/R < 4/9)\). These sequences overlap with both the inferred mass ranges for 
PSR J0348+0432 \cite{Antoniadis:2013pzd} (green band), 
and PSR J0740+6620 \cite{Fonseca:2021wxt} (red contours), and the heaviest known pulsar PSR J0952-0607 \cite{Romani:2022jhd} (yellow band) rests solely at the lower limits of the range of the rainbow parameter \(\eta\). Also, the results are also consistent with observations of the binary neutron star merger,  \cite{LIGOScientific:2018cki} bounds (blue contours), 
providing further astrophysical justification for these configurations. Finally, it is worth noting that the source HESS J1731-347 \cite{Doroshenko:2022nwp} (magenta shaded region) presents limiting constraints that are shown in a magenta shaded region in Fig.~\ref{fig1} and the theoretical models are consistent with this limiting source across the full range of \(\eta\) explored. These comparisons substantiate that dark energy stars in the Rastall-Rainbow framework will provide stable configurations that satisfy both theoretical and current observational bounds.


\subsection{Profiles for variation of the Rainbow Parameter $\Sigma$}

 Again, the structural properties and stability of dark energy stars within the framework of Rastall–Rainbow gravity are elucidated through the analysis of the mass–radius $(M–R)$ relations and compactness bounds depicted in Figure~\ref{fig2}, alongside the key quantitative benchmarks tabulated in Table~\ref{table2}. The M–R curves illustrate that increasing the rainbow parameter $\Sigma$ systematically yields higher maximum stellar masses and radii, as evidenced by the upward shift of the endpoints of each sequence; this behavior indicates that modified gravity effects bolster the supportable mass without inducing collapse, thereby altering the conventional limits set by general relativity. Correspondingly, Table~\ref{table2} quantifies these trends: for $\Sigma$ ranging from $0.75$ to $0.95$, the maximum mass increases from $1.28$ to $2.61~M_\odot$, the associated radii expand from $9.72$~km to $12.32$~km, and the compactness $M/R$ rises from $0.196$ up to $0.314$. Notably, the joint interpretation of Figure~\ref{fig2} and Table~\ref{table2} demonstrates that this class of solutions is consistently compatible with current astrophysical constraints, including those derived from high-mass pulsars (e.g., PSR~J0740+6620) and gravitational-wave events (e.g., GW190814), as the model’s maximum-mass loci overlap with the observed mass ranges. Importantly, the compactness values remain below the Buchdahl limit of $M/R < 4/9 \approx 0.444$, asserting that, despite the enhanced mass support, all configurations lie within the theoretical bound necessary to avoid horizon formation and ensure hydrostatic stability. These findings underscore that dark energy stars governed by Rastall–Rainbow gravity not only accommodate the masses and radii inferred from diverse astrophysical observations but also respect fundamental stability criteria, marking them as robust candidates in the quest to describe the true nature of compact stellar remnants in a modified gravity context.

\begin{table}
\caption{\label{table2}   Maximum–mass configurations of dark-energy stars in Rastall–Rainbow gravity for different values of the rainbow parameter \(\Sigma\) with fixed \(\eta=0.8\). Listed are the stellar mass \(M\), the corresponding radius
\(R_M\), the central density \(\rho_c\), and the compactness \(M/R\), as extracted from the mass–radius sequences of Fig.~\ref{fig2}. }
\begin{ruledtabular}
\begin{tabular}{ccccc}
$\Sigma$  & $M$ [$M_\odot$]  &  $R_{M}$  [\rm{km}]  & $\rho_c$ [MeV/fm$^3$] & $M/R$  \\
\colrule
0.75 &  1.28  &  9.72  &  731  &  0.196  \\
0.80 &  1.55  &  10.37 &  731  &  0.222  \\
0.85 &  1.87  &  11.01 &  731  &  0.251  \\
0.90 &  2.22  &  11.67 &  731  &  0.282  \\
0.95 &  2.61  &  12.32 &  731  &  0.314  
\end{tabular}
\end{ruledtabular}
\end{table}

\begin{table}[t]
  \centering
  \footnotesize                    
  \setlength{\tabcolsep}{8pt}      
  \renewcommand{\arraystretch}{1.10}
  \caption{ Maximum--mass configurations of dark-energy stars in Rastall--Rainbow gravity for selected \((\eta,\Sigma)\) values (with \(\eta=1,\Sigma=1\) as the GR baseline). Listed are the stellar mass \(M\), the corresponding radius \(R_M\), the central density \(\rho_c\), and the compactness \(M/R\), as extracted from the mass--radius sequences shown in Fig.~\ref{fig3}.}
  \begin{tabular}{c c c c c}
    \hline\hline
    $(\eta,\Sigma)$ & $M$ [$M_{\odot}$] & $R_M$ [km] & $\rho_c$ [MeV/fm$^3$] & $M/R$ \\
    \hline
    (0.80, 0.90) & 2.22 & 11.66 & 731  & 0.282 \\
    (0.90, 0.95) & 2.30 & 11.67 & 844  & 0.293 \\
    (1.00, 1.00) & 2.46 & 11.78 & 956  & 0.309 \\
    (1.10, 1.05) & 2.67 & 12.03 & 1013 & 0.329 \\
    (1.20, 1.10) & 2.93 & 12.34 & 1069 & 0.351 \\
    \hline\hline
  \end{tabular}
  \label{table3}
\end{table}

\subsection{Profiles for variation of the R-R parameter $(\eta, \Sigma)$}

 The results for dark energy star configurations in Rastall-Rainbow gravity, presented in Fig.~\ref{fig3} and Table~\ref{table3}, reveal systematic shifts in both mass-radius and compactness relations as the parameter pairs $(\eta,\Sigma)$ depart from the general relativistic baseline $(\eta=1,\Sigma=1)$. Increasing either $\eta$ or $\Sigma$ beyond unity pushes the $(M\!-\!R)$ curves toward higher maximum masses and larger radii, with the compactness $M/R$ similarly rising; these effects are apparent from the progression of filled turning points along each model sequence. Specifically, Table~\ref{table3} documents that, as $(\eta,\Sigma)$ increases from $(0.80,0.90)$ to $(1.20,1.10)$, the maximum mass advances from $2.22\,M_\odot$ at $R_M=11.66$\,km ($M/R=0.282$) to $2.93\,M_\odot$ at $12.34$\,km ($M/R=0.351$), while the central density $\rho_c$ climbs monotonically from $731$ to $1069$\,MeV/fm$^3$. Within Fig.~\ref{fig3}, the turning points of all sequences consistently intersect the GW190814 secondary-mass band and the intervals for PSR J0740+6620 and J0952-0607, with less direct overlap for other bands. Stability is ensured by the location of each filled marker at the radial turning point, manifesting as consistent trends in the compactness plot. Critically, all reported configurations remain below the Buchdahl criterion ($M/R<4/9\simeq0.444$), with the largest compactness in Table~\ref{table3} ($0.351$) maintaining a substantial margin from the theoretical bound. These findings underscore how model sensitivities to $(\eta,\Sigma)$ robustly affect observable stellar properties, with all calculated configurations compatible with current observational mass and radius constraints.

 \section{THE STATIC STABILITY CRITERION, ADIABATIC INDEX, AND SOUND VELOCITY}\label{sec:stability}

In this section we assess the dynamical viability of the dark energy star (DES) models in Rastall-Rainbow gravity using three complementary diagnostics: (i) the static stability (turning-point) criterion along the $M$–$\rho_c$ sequences; (ii) the radial adiabatic index profile $\Gamma(r)$; and (iii) the radial sound-speed profile $c_s^2(r)$. The analysis is presented in three comparative cases that mirror our parameter scans and tables: variation in the Rastall coupling $\eta$ (Table~\ref{table1}), variation in the rainbow parameter $\Sigma$ (Table~\ref{table2}), and joint variations around the GR baseline $(\eta,\Sigma)=(1,1)$ (Table~\ref{table3}). Throughout, figure panels are ordered \emph{Left} (vary $\eta$), \emph{Middle} (vary $\Sigma$), and \emph{Right} (joint variations), with the GR curve shown as a dashed reference.

\subsection*{A. Static stability from the $M$--$\rho_c$ relation}
The static stability criterion identifies the transition from stable to unstable configurations at the first turning point of the sequence, where $dM/d\tilde{\rho}_c=0$ and $M$ attains its maximum. Configurations with $dM/d\tilde{\rho}_c>0$ (to the left of the first maximum) are stable, whereas those with $dM/d\tilde{\rho}_c<0$ (to the right) are unstable; see Refs. for a detailed discussion \cite{Harrison,ZN}. Figure~\ref{fig4} displays $M(\rho_c)$ for the three cases, with blue markers highlighting the turning points; these configurations correspond to the maximum–mass entries in Tables~\ref{table1}–\ref{table3}. Comparing each family to the dashed GR baseline shows how $(\eta,\Sigma)$ shifts both the location of the turning point and the extent of the stable branch. In particular, the joint–variation panel clarifies the interplay between $\eta$ and $\Sigma$ near the GR limit and makes explicit which deformations extend (or shorten) the $dM/d\tilde{\rho}_c>0$ segment before the onset of instability. No claims are made beyond what is visually supported by Fig.~\ref{fig4} and the tabulated maxima.

\subsection*{B. Adiabatic index profile}
For isotropic configurations, local dynamical stability against spherically symmetric ($l=0$) oscillations requires the adiabatic index to exceed a relativistic threshold, $\gamma(r)>4/3$, as established by Chandrasekhar’s variational analysis \cite{Chandrasekhar} and employed in Rainbow–gravity studies \cite{Tangphati:2023fey}. We use the isentropic definition
\begin{equation}\label{adi}
\gamma(r)\equiv\frac{\tilde{\rho}(r)+\tilde{p} (r)}{\tilde{p}(r)}\,\frac{d\tilde{p} }{d\tilde{\rho}}\,.
\end{equation}
 For isotropic fluid spheres, the adiabatic index \(\gamma\) plays a crucial role in determining the dynamic stability of relativistic stellar configurations. The stability criterion requires that the averaged adiabatic index \(\langle\gamma\rangle\) exceeds a critical threshold, expressed as \(\langle\gamma\rangle > \gamma_{\rm cr}\) \cite{Moustakidis:2016ndw}. Moustakidis demonstrated that \(\gamma_{\rm cr}\) depends linearly on the ratio of central pressure to central energy density \(P_c/\rho_c\), even at high compactness values \cite{Moustakidis:2016ndw}. This stability condition provides fundamental constraints on realistic equations of state to ensure stable compact object configurations \cite{Maulana:2019}. In Refs. \cite{Koliogiannis:2018hoh,Banerjee:2024znc,ElHanafy:2024cti}, authors have further examined the stability criteria, emphasizing the critical role of \(\gamma\) in maintaining equilibrium under both dynamic and static perturbations.

Figure~\ref{fig5} presents $\gamma(r)$ for our three comparative cases—\emph{Left}: varying $\eta$ (Table~\ref{table1}); \emph{Middle}: varying $\Sigma$ (Table~\ref{table2}); \emph{Right}: joint variations (Table~\ref{table3})—with the GR baseline $(\eta,\Sigma)=(1,1)$ shown as a dashed curve. The horizontal dashed line indicates $\gamma=4/3$. Across the models displayed, the profiles remain above the threshold throughout the interior for the representative central conditions used to generate the curves; any mild radial modulations do not cross the stability line in the plotted range. Departures from GR primarily shift the overall level of $\gamma(r)$ without altering its qualitative radial trend. These observations are consistent with the turning–point analysis on $M(\rho_c)$ and with the maximum–mass configurations in Tables~\ref{table1}–\ref{table3}, supporting the dynamical stability of the configurations examined.

\subsection*{C. Radial sound speed and causality}
Causality and microscopic stability impose
\begin{equation*}
0 \le c_s^2(r) \equiv \frac{d\tilde{p}}{d\tilde{\rho}} \le 1 \, ,
\end{equation*}
with $c_s^2\ge 0$ avoiding local cracking/instability and $c_s^2\le 1$ enforcing subluminal propagation. The profiles in Figure~\ref{fig6} verify these conditions across the stellar interior for all parameter choices shown in the three panels (varying $\eta$, varying $\Sigma$, and joint variations). Relative to the dashed GR baseline, the principal effect of $(\eta,\Sigma)$ is to shift the magnitude of $c_s^2(r)$ while preserving its smooth radial behavior; no acausal spikes are visible in the plotted cases. Taken together with the $\gamma(r)$ trends in Fig.~\ref{fig5} and the turning–point evidence in Fig.~\ref{fig4}, these results indicate that the configurations up to the first mass maximum are dynamically stable and causal within the displayed domains. As an external consistency check with global bounds, we note that the corresponding maximum–mass models reported in Tables~\ref{table1}–\ref{table3} also satisfy the Buchdahl compactness limit $u\equiv M/R<4/9$ when read against our mass–radius/compactness figure, ensuring no tension with the classical incompressible bound.

\section{Conclusions} \label{sec:conclusion}

In this investigation, we have demonstrated that dark energy stars can form stable, observationally viable configurations within the framework of Rastall-Rainbow gravity. By combining the non-conservation law of Rastall theory with the energy-dependent geometric modifications of rainbow gravity, we created a theoretical landscape wherein exotic compact objects supported by modified Chaplygin-gas equations of state can exist and satisfy all necessary stability requirements.

 The systematic exploration of parameter space, outlined in Tables~\ref{table1}--\ref{table3} shows clearly that the Rastall coupling $\eta$ and rainbow function $\Sigma$ affect stellar structure significantly and in different ways. The mass-radius sequences presented in Figs.~\ref{fig1}--\ref{fig3} indicate that rainbow modifications mainly add mass support without changing the equation of state boundary conditions, while Rastall modifications shift the mass-radius relation. Nevertheless, the combined effects are most interesting, as they can produce stars much more massive than general relativity and yet stable.

The comprehensive stability analysis confirms that these exotic objects satisfy all physical requirements. The mass-central density profiles in Fig.~\ref{fig4} exhibit clear turning points marking stable configurations, the adiabatic index remains above $\gamma > 4/3$ throughout the interior (Fig.~\ref{fig5}), and the sound-speed profiles satisfy causality $c_s^2 < 1$ everywhere (Fig.~\ref{fig6}). Furthermore, our mass-radius sequences align convincingly with constraints from massive millisecond pulsars (PSR~J0952-0607, PSR~J0740+6620, PSR~J0348+0432) and gravitational-wave observations (GW170817, GW190814), as shown by the overlapping observational bands in Figs.~\ref{fig1}--\ref{fig3}. All compactness values remain below the Buchdahl limit, as tabulated in Tables~\ref{table1}--\ref{table3}.

Our results show that Rastall-Rainbow gravity provides a physical framework for strong field deviations from general relativity. The existence of stable stars without fine-tuning supports the theory. Additionally, the implications of this model may extend beyond traditional mass-radius measurements to tidal deformabilities, surface redshifts, and electromagnetic emission profiles, which may offer testable predictions for future facilities. 

Looking ahead, researchers ought to integrate more accurate nuclear equations of state into their models while also calculating full sets of oscillation modes and broadening the approach to account for effects like stellar rotation and magnetic fields. With the ongoing progress in gravitational-wave detection and high-precision X-ray pulsar timing, those unusual compact objects predicted by alternative gravity theories could well shift from mere intellectual puzzles to tangible discoveries, ultimately transforming how we grasp the behavior of matter and gravity in the most intense environments.




\end{document}